# Soft x-ray photoemission study of new BiS$_2$-layered superconductor LaO$_{1-x}$F$_x$BiS$_2$

Shinsuke Nagira,[1] Junki Sonoyama,[1] Takanori Wakita,[1,2] Masanori Sunagawa,[1]
Yudai Izumi,[3] Takayuki Muro,[3] Hiroshi Kumigashira,[4] Masaharu Oshima,[5,*]
Keita Deguchi,[6] Hiroyuki Okazaki,[6] Yoshihiko Takano,[6] Osuke Miura,[7] Yoshikazu Mizuguchi,[7]
Katsuhiro Suzuki,[8] Hidetomo Usui,[9] Kazuhiko Kuroki,[9]
Kozo Okada,[10] Yuji Muraoka,[1,2] and Takayoshi Yokoya[1,2]

[1]Research Laboratory for Surface Science, Okayama University, Okayama 700-8530, Japan
[2]Research Center of New Functional Materials for Energy Production, Storage, and Transport, Okayama University, Okayama 700-8530, Japan
[3]Japan Synchrotron Radiation Research Institute (JASRI) /SPring-8, 1-1-1 Kouto, Sayo, Hyogo 679-5198, Japan
[4]High Energy Accelerator Research Organization (KEK), Photon Factory, Tsukuba, Ibaraki 305-0801, Japan
[5]Department of Applied Chemistry, The University of Tokyo, Tokyo 113-8656, Japan
[6]National Institute for Materials Science, 1-2-1, Sengen, Tsukuba, 305-0047, Japan
[7]Department of Electrical and Electronic Engineering, Tokyo Metropolitan University, 1-1, Minami-osawa, Hachioji, 192-0397, Japan
[8]Department of Engineering Science, The University of Electro-Communications, Chofu, Tokyo 182-8585, Japan
[9]Department of Physics, Osaka University, 1-1 Machikaneyama, Toyonaka, Osaka 560-0043, Japan
[10]Department of Physics, Okayama University, Okayama 700-8530, Japan
*Present address: Synchrotron Radiation Research Organization, The University of Tokyo, Tokyo 113-8656, Japan

**Abstract**

We use core level and valence band soft x-ray photoemission spectroscopy (SXPES) to investigate electronic structure of new BiS$_2$ layered superconductor LaO$_{1-x}$F$_x$BiS$_2$. Core level spectra of doped samples show a new spectral feature at the lower binding energy side of the Bi 4$f$ main peak, which may be explained by core-hole screening with metallic states near the Fermi level ($E_F$). Experimental electronic structure and its $x$ dependence (higher binding energy shift of the valence band as well as appearance of new states near $E_F$ having dominant Bi $6p$ character) were found to be consistent with the predictions of band structure calculations in general. Noticeable deviation of the spectral shape of the states near $E_F$ from that of calculations might give insight into the interesting physical properties. These results provide first experimental electronic structure of the new $BiS_2$ layered superconductors.

Recent discovery of a new $BiS_2$-layered superconductor having a $BiS_2$ layer as a common crystal structure [1-9] has provoked attention because of the interesting normal state properties [3,4,6,9,10,11] and relation of the superconductivity to charge density wave (CDW) [12,13], driven by quasi one dimensional Fermi surface sheets [14]. The highest superconducting critical temperature ($T_c$) of 10.6 K has been reported for $LaO_{1-x}F_xBiS_2$ [3], which consists of double $BiS_2$ layers sandwiched between $La_2O_2$ spacer layers (Fig. 1). The parent compound shows semiconducting temperature dependence in resistivity. As fluorine concentration ($x$) is increased, superconductivity appears at $x = 0.2$, and the value of $T_c$ increases till $x = 0.5$ and decreases for further $x$ [9]. Interestingly, the compounds showing superconductivity exhibit semiconducting behavior [3,4,6,9], similar to under-doped region of high $T_c$ compounds [15], for which competition and/or coexistence of itinerant electrons with ordered phases have been discussed. Thus $BiS_2$ layered new superconductors might give us another opportunity of studying superconductivity competing and/or coexisting with ordered phases.

Band structure calculations have reported that $LaOBiS_2$ is an insulator having a band gap of ~ 0.8 eV [12,13,16,17]. While the valence band consists of S(S1 and S2, see Fig. 1) $3p$ and O $2p$ orbitals, conduction band is derived from Bi $6p_{x,y}$ and S1 $3p$. Since the O substitution of F introduces electron carriers, the chemical potential of $LaO_{0.5}F_{0.5}BiS_2$ is expected to locate within the Bi $6p_{x,y}$ and S1 $3p$ hybridized band with substantial dispersion. Location of the chemical potential at the highly dispersive bands makes the system normal metal as a rule. But, this simple expectation does not hold, as the doped samples show the semiconducting behavior. In relation to this, band structure calculations have predicted strong nested Fermi surface having quasi one-dimensional nature [14] and presence of CDW instability for $x = 0.5$ [12,13], which can explain pressure dependence of $T_c$ in $LaO_{0.5}F_{0.5}BiS_2$ [11]. For $x = 0.0$, highly anharmonic ferroelectric soft phonons at the zone center have been predicted [13]. Moreover, possibilities of exotic paring mechanisms due to electron-electron correlation originating from the nested Fermi surface have also been discussed [14,18-20]. Therefore in order to understand the mechanism of the superconductivity, it is highly desirable to experimentally investigate the normal state electronic structure that leads to the superconductivity.

In this paper, we report core level and valence band electronic structure of $LaO_{1-x}F_xBiS_2$ measured using soft x-ray photoemission spectroscopy (SXPES). SXPES gives spectra that reflect bulk valence band density of states (DOS) and core levels, and is therefore a suitable technique to investigate electronic structure and chemical state of solids. $x$ dependent core level SXPES shows systematic core level shifts as well as emergence of a new spectral component in the Bi $4f$ core level. $x$ dependent valence band SXPES shows a shift of the valence band and appearance of new states at the Fermi level ($E_F$) with substantial Bi $6p$ character, consistent with band structure calculations in general. On the other hand, noticeable deviation of the intensity and spectral shape of the states near $E_F$ from those of calculations was observed, latter of which might have relation to the interesting physical properties.

Polycrystalline samples of $LaO_{1-x}F_xBiS_2$ were prepared using powders of $Bi_2O_3$ (98%), $BiF_3$ (99.9%), $La_2S_3$ (99.9%), $Bi_2S_3$ and Bi (99.9%) grains. The $Bi_2S_3$ powder was prepared by a reaction between Bi grains and S (99.9999%) grains at 500 °C in an evacuated quartz tube. The starting materials with a nominal composition of $LaO_{1-x}F_xBiS_2$ ($x$ = 0.0, 0.3, and 0.5) were mixed, pressed into pellets, sealed into an evacuated quartz tube, and heated at 700 °C for 10 h. The product was reground, pressed into pellets, and annealed a second time at 600 °C under a high pressure of 2 GPa using a cubic-anvil high-pressure synthesis instrument with 180 ton press. From temperature dependence of magnetic susceptibility, the onsets of $T_c$ of the samples were estimated at values of 9.8 K and 10.4 K for $x$ = 0.3 and 0.5, respectively. For $x$ = 0.0, we observed a non diamagnetic signal above 2 K.

Core level and valence band SXPES measurements were performed at BL27SU in SPring-8, with a PHOIBOS 150 electron analyzer. SXPES measurements near $E_F$ were performed at BL2C in Photon Factory (PF), KEK, with a SES2000 electron analyzer. The total energy resolution was set to 0.2 eV and 0.1 eV for PES measurements at SPring-8 and PF, respectively. The binding energies were referenced to $E_F$ of gold. The base pressures of the measurement chambers were better than $5.0 \times 10^{-8}$ Pa. The samples were fractured to get fresh surfaces at 300 K and kept at the same temperature during the measurements in order to prevent possible charging up effect. As we used polycrystalline samples, obtained photoemission (PE) spectra reflect the integrated DOS. All PE results reported here have been obtained within 6 h after fracturing and no spectral changes were observed during the measurements, indicating that the spectra shown here represent intrinsic electronic structures of the compounds.

To compare the experimental valence band spectrum with theoretical studies, we also calculated the band structure of $LaO_{0.5}F_{0.5}BiS_2$ using the Wien2k package [21] and adopting the lattice structure given in Ref. 3. We adopt the virtual crystal approximation (VCA) in order to take into account the effect of the element substitution. In VCA, an averaged mixture of oxygen and fluorine potentials is considered, so that the description of those states that directly originate from these atoms may not be quantitatively reliable. Since these states appear in the higher binding energy region of the valence band, we compare the calculation with SXPES data only for the lower binding energy region where S and Bi derived states predominate.

Figure 2 shows $x$ dependent Bi $4f$ and S $2p$ core level spectra of LaO$_{1-x}$F$_x$BiS$_2$ (open circles) measured with a 500 eV photon energy, together with results of curve fitting analyses (lines). Each spectrum consists of four peaks: two dominant peaks around 158.5 eV and 164 eV are spin-orbit partners of Bi $4f$ and the two smaller peaks between them are those of S $2p$. The four peaks shift toward higher binding energy as a function of $x$, consistent with electron doping due to substitution of oxygen with fluorine. Surprisingly, the spectral shape of Bi $4f$ core level drastically changes with F substitution; the intensity at the lower binding energy side of Bi $4f_{7/2}$ of $x$ = 0.3 and 0.5 is higher than that of $x$ = 0.0. In order to confirm the existence of the new feature, we have performed core level analyses using four Voigt functions for four peaks of the Bi $4f$ and S $2p$ core levels. We added two components around 157 eV and 159 eV and their spin-orbit counterparts, which can be ascribed to components due to Bi metal and Bi$_2$S$_3$/Bi$_2$O$_3$, respectively [22]. We also used another component II to reproduce the new feature of $x$ = 0.3 and 0.5. Other than the known spectral weight ratios and energy differences between the spin-orbit components of $p$ and $f$ orbitals, no further restrictions were made for final convergence. As shown in Fig. 2, the spectra of $x$ = 0.3 and 0.5 are explained with addition of new component II. These results indicate existence of new spectral component of Bi induced by F substitution, which is unexpected from only one crystallographic site for Bi in LaO$_{1-x}$F$_x$BiS$_2$.

One of the explanations for the multiple components may be mixed valence, as observed for Fe$_3$O$_4$ [23]. For BaBiO$_3$, the parent compound of Ba$_{1-x}$K$_x$BiO$_3$ and Ba$_{1-x}$Pb$_x$BiO$_3$, it was suggested that mixed valence of Bi 3+ and Bi 5+, instead of Bi 4+, was realized [24], though x-ray absorption spectroscopy measurements at the Bi L$_3$ edge has proposed a Bi 4+ valence [25]. From an ionic picture, the formal valence of Bi in LaOBiS$_2$ is 3+ and that in LaO$_{0.5}$F$_{0.5}$BiS$_2$ is 2.5+, assuming the formal valence of S of 2-. The change of the spectral shape of Bi $4f$ core level can be thought of increasing Bi 2+ component in the mixed valence of Bi 3 + and 2 +, possibly fluctuating with a time scale slower than that of PES measurements (10$^{-15}$ s) [26] in the superconductive samples. However, the Bi valence of 2+ is rare in real materials, which is different from BaBiO$_3$ where Bi 3+ and Bi 5+ are chemically more stable than the formal valence of Bi 4+. A more plausible explanation may be related to core-hole screening process. In electron-doped cuprate high-$T_c$, Nd$_{2-x}$Ce$_x$CuO$_4$ [27], emergence of the doping-induced new core level component at a lower binding energy side of the main peak was explained with a core-hole screening by doping-induced states at $E_F$. For $LaO_{0.5}F_{0.5}BiS_2$, the new feature can be ascribed to the $4f^{13}6p^1\underline{C}$ state, where the $6p^1\underline{C}$ represents the charge transfer between Bi $6p$ state and the doping-induced state at $E_F$. In this case, energy difference between two components (main and lower binding energy feature) is known to correspond to the width of band gap. This agrees with the fact that the energy difference of ~ 0.7 eV between components I and II estimated from the core level analyses for $LaO_{1-x}F_xBiS_2$ ($x$ = 0.3 and 0.5) is similar to the calculated width the band gap (~ 0.7 eV). F-induced change in Bi core level spectral shape provides spectroscopic evidence that the doped electrons enter into Bi $6p$ orbitals.

Figure 3 shows the valence band SXPES spectra of $LaO_{1-x}F_xBiS_2$ ($x$ = 0.0, 0.3, and 0.5) obtained using 500 eV (open circles connected with lines) photons, together with that of $LaO_{0.5}F_{0.5}BiS_2$ with 1100 eV (a broken line) photons. The valence band spectrum of $LaOBiS_2$ ($x$ = 0.0) has a peak around 2.5 eV with a structure around 5 eV. The spectrum shows a negligible intensity region from $E_F$ to 1 eV binding energy, which indicates an experimental band gap of 1eV below $E_F$. As $x$ is increased, the whole valence band shifts to higher binding energy by ~ 0.3 eV, as is seen from the shift of the spectral edge of the valence band around 2 eV. The $x$ dependent shift to the higher binding energy side is consistent with electron doping by substitution of oxygen with fluorine. For the spectral shape, the structure around 5 eV gets broader and a new peak evolves around 8.5 eV, latter of which also shifts to higher binding energy. Absence of the peak around 8.5 eV in the $x$ = 0.0 sample and its systematic evolution indicate that the structure is fluorine-derived states. Importantly, it is observed that the intensity near $E_F$ increases for higher $x$.

According to band calculations [12,13,15,16], $LaOBiS_2$ has a valence band derived from S(S1 and S2) $3p$ and O $2p$ orbitals with a ~ 4 eV band width, which is separated by a band gap of ~ 0.8 eV from the bottom of the highly dispersive conduction band of dominant Bi $6p_{x,y}$ character hybridized with S1 $3p$. Introduced electrons by substitution of O with F fill the conduction band and, for $x$ = 0.5, $E_F$ is expected to be located at ~ 0.8 eV above the bottom of the conduction band. For the $x$ = 0.0 sample, the fact that the experimental band gap of 1 eV below $E_F$ is comparable with that of the calculated band gap suggests that $E_F$ may be located near the bottom of the conduction band possibly due to impurity and/or self doping [6]. Assuming that $E_F$ of $x$ = 0.0 sample is located at the bottom of the conduction band, we may deduce the chemical potential shift of ~ 0.3 eV between $x$ = 0.0 and 0.5 samples. This value is much smaller than that expected from band calculations (0.8 eV).

Corresponding the observation with band structure calculations, we may ascribe that the main valence band is dominantly derived from the S $3p$ and O $2p$. The bottom of the conduction band is mainly derived from the Bi $6p_{x,y}$ and S1 $3p$. The peak around 8.5 eV is ascribed to states derived from F $2p$. The orbital character of the states near $E_F$ for the $x = 0.5$ sample, which is essential in understanding the physical properties including superconductivity, can be verified with photon-energy dependence of the photoionization cross section for different orbitals [28]. In Fig. 3, we compared the valence band spectrum of a 1100eV photon energy (a broken curve in Fig. 3) with the spectrum of a 500 eV photon energy. The normalization of the 1100 eV spectrum to the 500 eV one was performed with the intensity of the structure around 3 eV, which has dominant S $3p$ character. From 500 eV to 1100 eV, the ratio of photoionization cross sections of Bi $6p$ to S $3p$ increases. Near $E_F$ region (~1.0 eV) in Fig. 3, the intensity of 1100 eV is enhanced by the factor of about 2 compared to that of 500 eV, which agrees well with band calculations predicting dominant Bi $6p$ character of the states at $E_F$ over S $3p$.

On the other hand, we noticed that the intensity as well as the spectral shape near $E_F$ of $x = 0.5$ eV deviates from band structure calculations. In Fig. 3, we plotted a theoretical photoemission spectrum of $x = 0.5$, which was obtained by Gaussian and energy-dependent Lorentzian broadening of the sum of the calculated partial DOS after taking photoionization cross sections of 500 eV [28] into account. The calculated spectrum was normalized to the intensity around 3 eV. The intensity near $E_F$ of the experimental spectrum is smaller than that of the theoretical spectrum. This may correspond to the observation that the shift of the valence band is smaller than that of calculation, which suggests that the actual electron doping level may be smaller than that estimated from nominal F concentration. For the spectral shape near $E_F$ (Fig. 4), we found that the experimental spectrum appears qualitatively different from the calculation, which cannot be explained with the possible smaller doping level alone. While spectrum shows a spectral edge similar to that of Au indicative of metallic character, it shows a nearly flat intensity distribution near $E_F$, in contrast to the increasing intensity toward $E_F$ of the calculated spectrum. In addition, a smoothed symmetrized spectrum has a broad hump around 0.3 eV. Such an anomalous spectral behavior with reduced DOS near $E_F$ compared to the calculated DOS is reminiscent of incoherent spectral weight observed in correlated materials [29] and might give a hint to understand the semiconducting-like resistivity [3,4,6,9] and its relation to the predicted quasi-nested Fermi surface leading to the CDW instability [12-14]. Higher-resolution PES study, preferably ARPES, using single crystal samples is valuable in experimentally establishing the anomalous spectral shape and will give a deeper insight into the interesting physical properties of the new superconductors.

Using core level and valence band SXPES, we investigated electronic structure of new BiS$_2$ layered superconductor LaO$_{1-x}$F$_x$BiS$_2$. Core level spectrum shows appearance of a new spectral component at lower binding energy sides of Bi 4$f$, which may be explained by core hole screening with metallic states near $E_F$. Valence band SXPES results give electronic structure information on the new superconductor, including the shape of the valence band, $x$ dependent energy shift, and appearance of states at $E_F$ with Bi 6$p$ character, which show overall agreement with band structure calculations. On the other hand, marked deviation of the spectral shape of the states near $E_F$ from those of calculation may give a hint to understand the anomalous physical properties.

Acknowledgments
We thank K. Machida for valuable discussion. We thank T. Otsuka for help in PES measurements at SPring-8. PES measurements at BL27SU of SPring-8 were performed under a proposal number 2012B1693. PES measurements at BL2C of PF were performed under a project 2011S2-003.

Figure Captions

Fig. 1(color online)    Crystal structure of $LaO_{1-x}F_xBiS_2$

Fig. 2(color online)    Bi 4$f$ and S 2$p$ core level SXPES spectra of $LaO_{1-x}F_xBiS_2$ ($x$ = 0.0, 0.3, and 0.5) (open circles) and the results of curve fitting (curves). Normalization for $x$-dependent spectra was performed with the intensity of Bi 4$f_{7/2}$.

Fig. 3(color online)    Valence band SXPES spectra of $LaO_{1-x}F_xBiS_2$ ($x$ = 0.0, 0.3, and 0.5) obtained using 500 eV (open circles connected with lines) and 1100 eV (a broken line) photon energies (only for $LaO_{0.5}F_{0.5}BiS_2$). All the spectra measured with a 500 eV photon energy were normalized to the intensity of the peak around 2 - 3 eV binding energy. The normalization of the 1100 eV spectrum to the 500 eV one for the $x$ = 0.5 sample was performed at the intensity of the peak around 3 eV. A calculated spectrum for a 500 eV photon energy made from calculated DOS for $x$ = 0.5 is also shown, for which energy resolution, life time broadening, and Fermi Dirac distribution function at 300 K were taken into account. Note that the effect of using an imaginary atom in VCA approximation appears in the higher binding energy region of the valence band in this compound, as described in the text.

Fig. 4(color online)　SXPES spectrum near $E_F$ of $LaO_{0.5}F_{0.5}BiS_2$ measured with a 500 eV photon energy and with a higher energy resolution, which is compared with the calculated spectrum and the spectrum of Au. Normalization of the experimental and calculated spectra of $LaO_{0.5}F_{0.5}BiS_2$ was done with the intensity of the peak around 3eV, and the 1/2 intensity of the calculated spectrum is plotted. The smoothed symmetrized spectrum of the experimental spectrum of $LaO_{0.5}F_{0.5}BiS_2$ is also shown for emphasizing the difference in spectral shape between the experimental and calculated spectra.

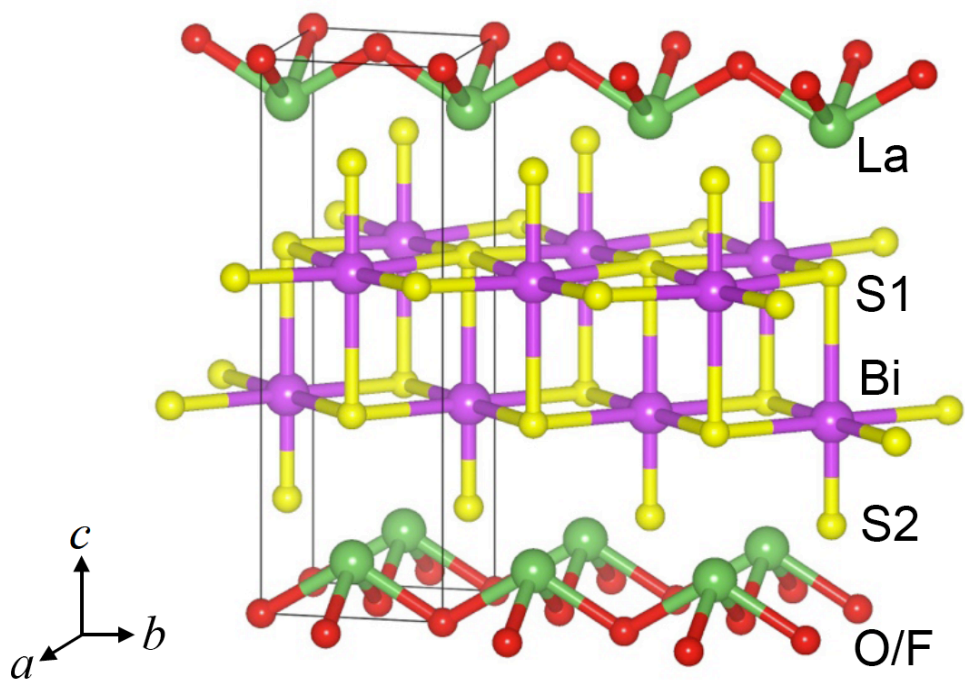

Fig. 1    S. Nagira

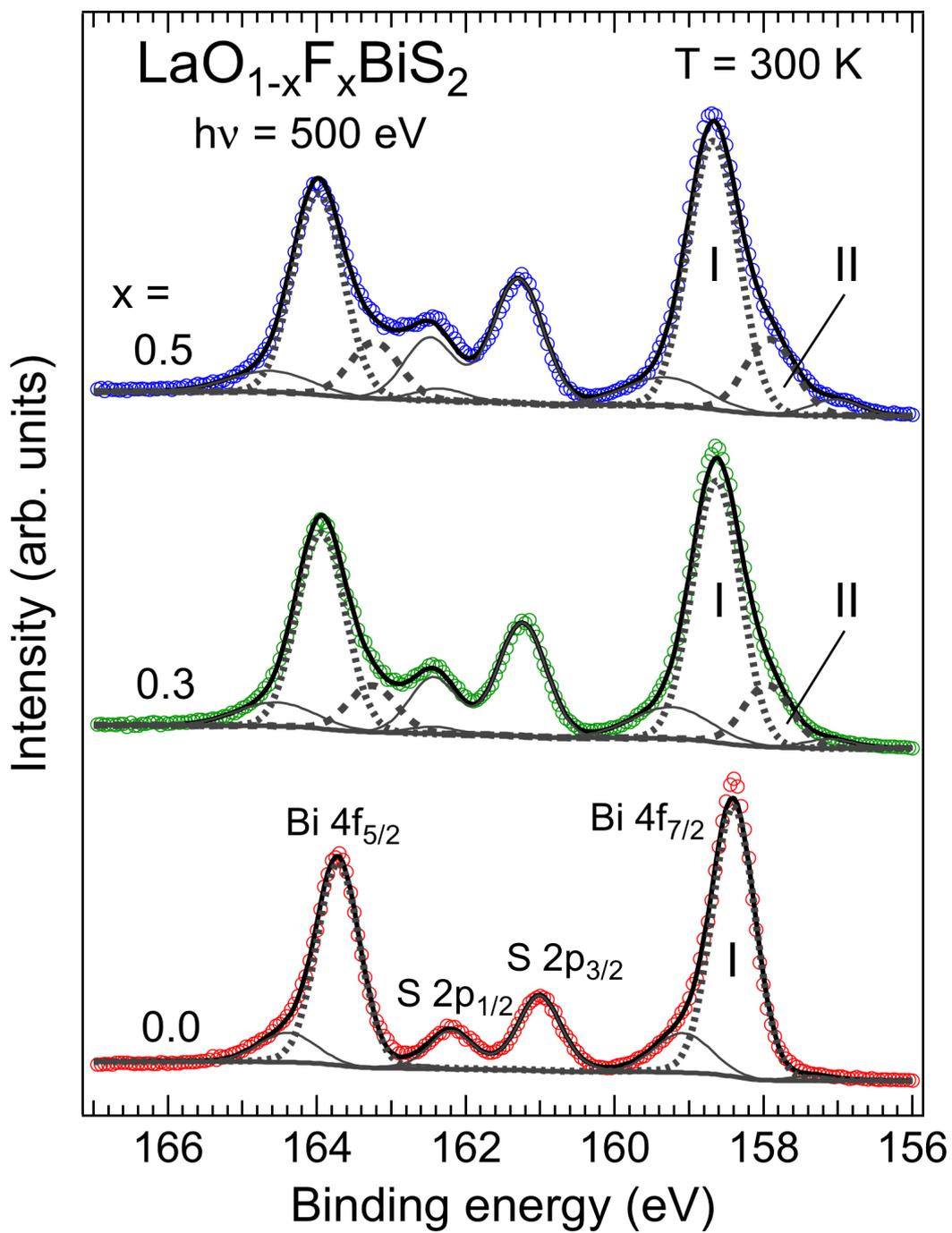

Fig. 2  S. Nagira

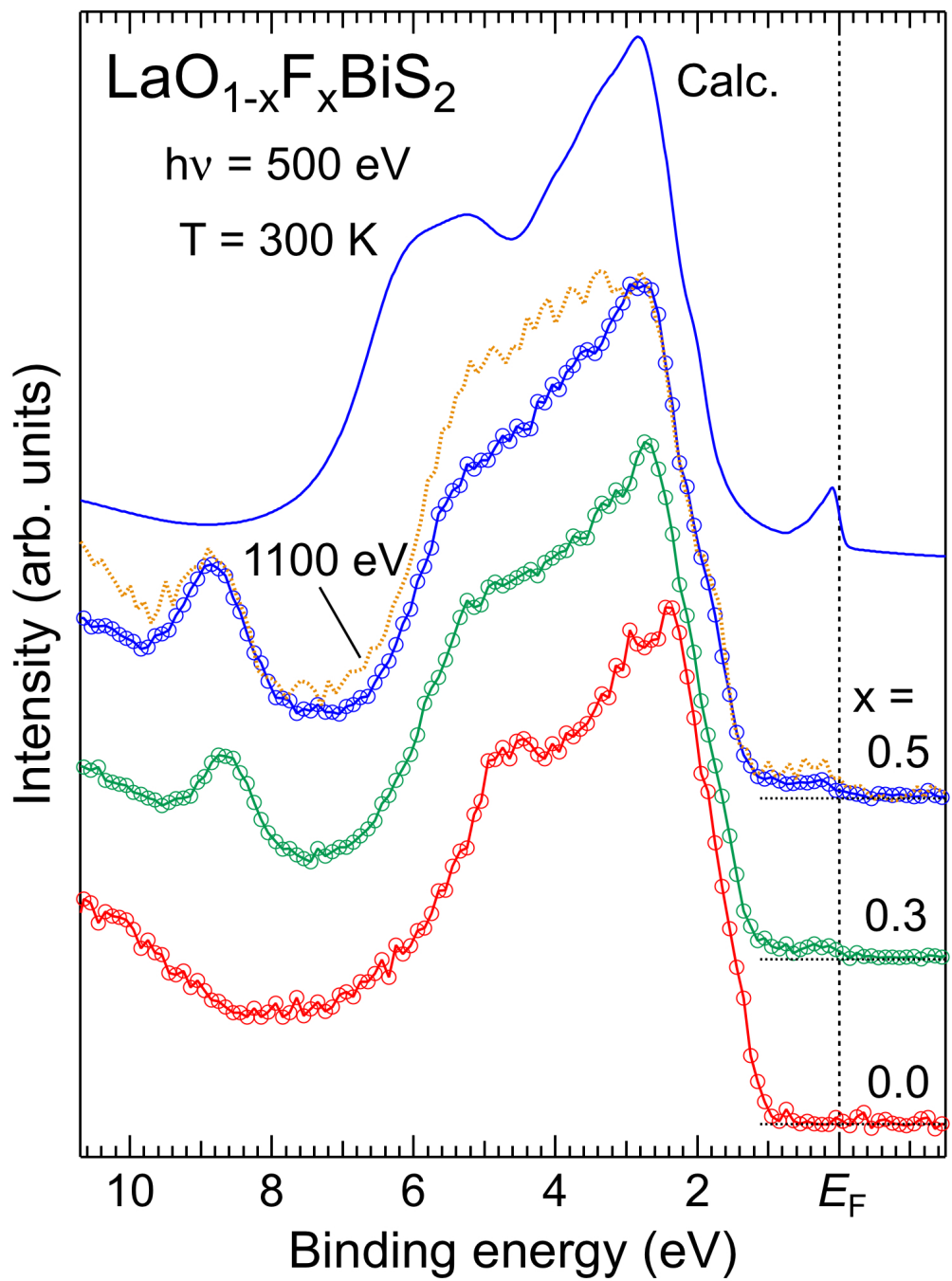

Fig. 3   S. Nagira

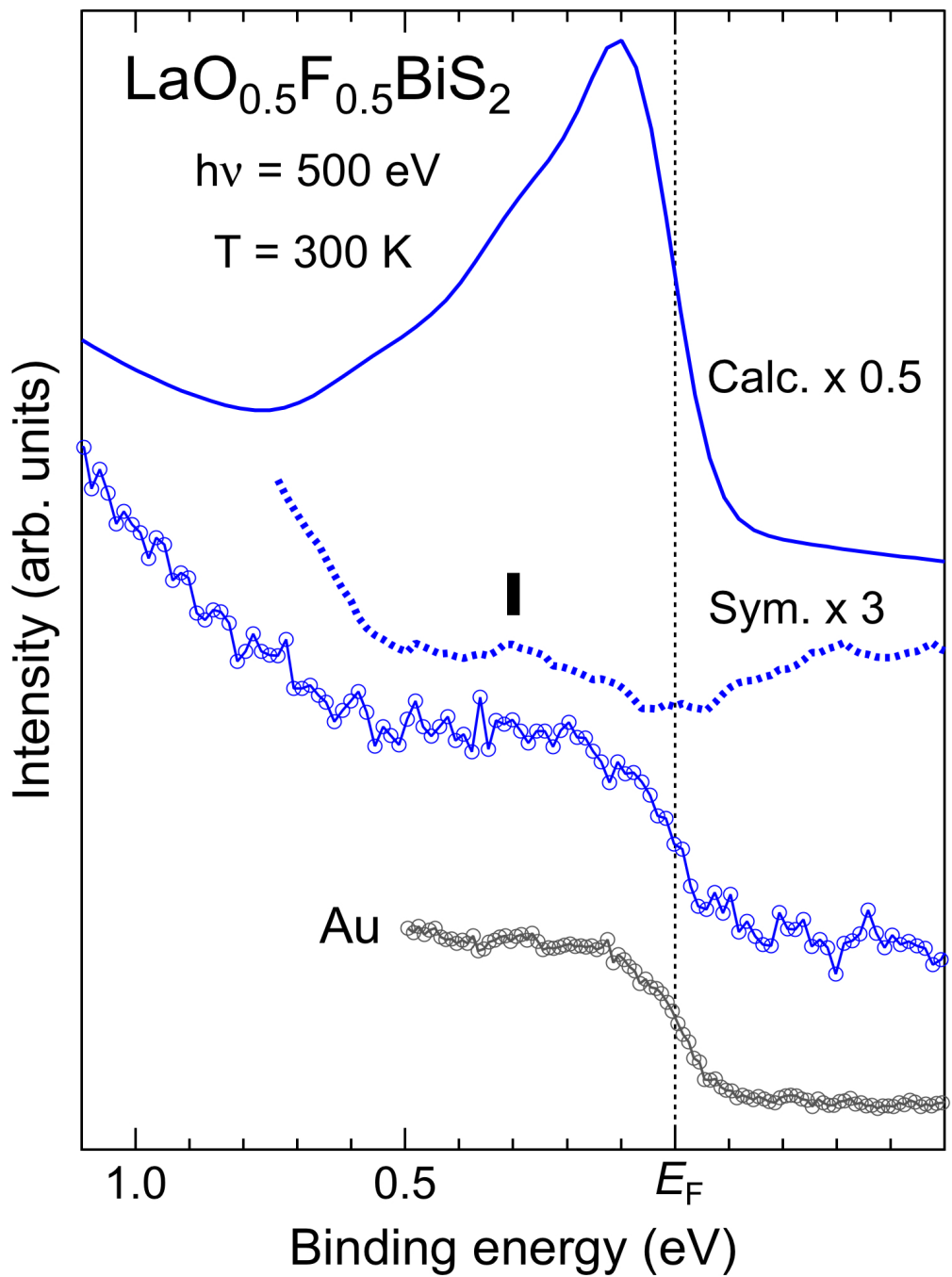

Fig. 4    S. Nagira